\documentclass[preprint,showpacs,preprintnumbers,amsmath,amssymb]{revtex4}

\usepackage{graphicx}
\usepackage{dcolumn}
\usepackage{bm}

\begin{document}

\preprint{APS/}

\title{Direct Measurement of Kirkwood-Rihaczek distribution for spatial properties of coherent light beam}

\author{Viktor Bollen, Yong Meng Sua and Kim Fook Lee}
\email[]{kflee@mtu.edu}
\affiliation{%
Department of Physics,\\ Michigan Technological
University,\\ Houghton, Michigan 49931}%

\date{\today}

\begin{abstract}
We present direct measurement of Kirkwood-Rihaczek (KR)distribution
for spatial properties of coherent light beam in terms of position
and momentum (angle) coordinates. We employ a two-local oscillator
(LO) balanced heterodyne detection (BHD) to simultaneously extract
distribution of transverse position and momentum of a light beam.
The two-LO BHD could measure KR distribution for any complex wave
field (including quantum mechanical wave function) without applying
tomography methods (inverse Radon transformation). Transformation of
KR distribution to Wigner, Glauber Sudarshan P- and Husimi or Q-
distributions in spatial coordinates are illustrated through
experimental data. The direct measurement of KR distribution could
provide local information of wave field, which is suitable for
studying particle properties of a quantum system. While Wigner
function is suitable for studying wave properties such as
interference, and hence provides nonlocal information of the wave
field. The method developed here can be used for exploring spatial
quantum state for quantum mapping and computing, optical phase space
imaging for biomedical applications.

\end{abstract}

\maketitle

\section{Introduction}

Optical phase-space
tomography~\cite{kim99,wax96,wax01,Reil05,walmsley96,cheng99,smith05,mukamel03}
is an optical method for characterizing spatial properties of light
fields/photons. The Kirkwood (K(x,p))~\cite{kirkwood33} and Wigner
(W(x,p))~\cite{wigner32} distribution functions were originally
proposed in the studies of quantum statistics and thermodynamic for
almost classical ensembles. The Kirkwood distribution
($\psi(x)\psi^{*}(p)e^{-ixp}$)has been rediscovered by
Rihaczek~\cite{rihaczek68} for use in the theory of time-frequency
analysis of classical signals. The Wigner function is more popular
than the KR function because it has some unique properties such as
negative values, real and symmetry. It can be obtained through
tomograph methods and applicable to problems in phase space
transport equation such as Liouville equation~\cite{sjohn96}. The
Wigner function was first introduced in optical light field by
Bastiaans~\cite{bastiaans78, bastiaans97} to analysis spatial
properties of an optical Gaussian beam. In quantum optics, the
quantum mechanical wave function cannot be measured in principle.
The wave function or quantum state of a physical system can be best
represented by the Wigner function. Vogel and Risken~\cite{vogel89}
had theoretically proposed how quadrature amplitudes of nonclassical
light fields can be represented in Wigner distribution by tomograph
methods.
Raymer~\cite{smithey93,lvovsky09,raymer94,mcalister95,beck93} has
pioneered the tomography measurement of Wigner function for
quadrature amplitudes of squeezed light, spatial properties and
time-frequency properties of coherent light (coherent state with
large mean photon number).

In most quantum information experiment, a single spatial mode of
electromagnetic field is generally used and
considered~\cite{kolobov06,kolobov99}. Quantum fluctuations of light
at different spatial points in the transverse plane of the light
beam have to be taken into account. Kolobov and
Sokolov~\cite{kolobov86,kolobov89,kolobov89b,kolobov89c,kolobov91}
have studied in detail the possibility of local squeezing in a
spatial mode of light beam. Multimode spatial modes would cause
information processing and computing errors. The property of
multimode squeezed light which allows us to increase the sensitivity
beyond the shot-noise limit, could create many interesting new
applications in optical imaging, high-precision optical
measurements, optical communications and optical information
processing. However, one has to employ a dense array of
photodetectors to observe multimode squeezed states.

The phase space distributions from the Glauber Cahill
s-parameterized class of quasi-distributions~\cite{cahil69} that
contain the Wigner function, the Glauber Sudarshan
P-representation~\cite{sudarshan63}, and the Husimi~\cite{husimi40}
or the Q-representation, have been widely used as powerful phase
space tools. However, most of tomograph methods are not involving
direct measurement of these quasi-quantum distribution functions.
These distributions are usually reconstructed from the raw data,
where numerical and mathematical transformation discrepancies may
have made these distributions doubtfully representing the true
states. The direct measurement of these distributions are desirable
to represent the true quantum state. It is well understood that
tomograph methods developed for these distributions can be used for
any complex wave fields regardless of classical or quantum
origin~\cite{smith07,lvovsky09}. In this perspective, we use wave
field to represent quantum mechanical wave function and coherent
light field.

The KR distribution for any quantum state in a generalized form has
been introduced~\cite{wodkiewicz03b}. The hydrogen atom has been
investigated using the Kirkwood$-$Rihaczek phase space
representation~\cite{wodkiewicz03}. It has been pointed out that the
KR distribution may not suitable for a "rotated state". The Wigner
function for a ''rotated state'' is just simply the rotated Wigner
function. This is the basis property in tomography method applied in
reconstruction of Wigner function. However, this simple relation
does not hold for KR function. Tomograph methods (Radon Transform)
are not suitable for KR function. It is worth noting that position
and momentum distributions of wave field can be obtained through the
KR distribution such as $\int K(x,p) dp = |\psi(x)|^2$ and $\int
K(x,p) dx = |\psi(p)|^2$, respectively. These properties are similar
to Wigner function. Moreover, the complex conjugated KR can be
directly measured through a two-LO technique~\cite{kim99}, not like
Wigner function which is reconstructed from experimental data using
quantum tomographic methods. Phase-space distribution functions
especially Wigner function can unravel unique quantum properties
such as entanglement of correlated
systems~\cite{schleich06,wodkiewicz99}, negative parts of
phase-space plots~\cite{schleich01}, and the phase space sub-Planck
structures of quantum interference~\cite{zurek06,zurek01}. The KR is
relatively unexplored for quantum fields.

Spatial coherence of light sources is necessary for achieving
efficient coupling into fiber systems for quantum communication, and
also for biomedical  imaging in image-guide
intervention~\cite{brezinski08}. From a quantum-mechanical
perspective, spatial degree of freedom of a photon is another
optical quantum realization for encoding information such as spatial
qubits~\cite{kolobov06}. Full characterization of arbitrary,
continuous spatial states of photons is important for understanding
the concept of the photon wave function~\cite{smith07}. Wigner
function for an ensemble of identically prepared photons in the
transverse spatial modes can be completely used to characterize the
transverse spatial state of the ensemble. The KR distributions are
relatively unexplored for spatial properties of wave fields. In the
second quantization of quantum mechanics, electric field is written
as operator in term of harmonic oscillator basis, that is, position
and momentum operators. The spatial property of the electric field
perpendicular to the propagation direction is described by mode
functions. The mode function in x or p coordinate is a description
of probability amplitude to find a photon at transverse position x
or transverse momentum p, respectively. The field operator
$\hat{a}(x)$ corresponding to the mode function will provide mean
field $\langle \hat{a}(x)\rangle$ and quantum noise $\triangle
\hat{a}(x)$ in an optical detection scheme. In this work,
distribution of x and p of a coherent wave field is measured by
using a two-LO balanced heterodyne detection. Mean value measurement
of the wave field $\langle \hat{a}(x)\rangle$ at a configuration
space will provide classical-like feature regardless the origin
(classical or quantum light) of the measured field. Variance
measurement of the wave field $\triangle \hat{a}(x)$ will provide
quantum feature of the light field. For coherent state with low mean
photon number or large mean photon number, the variance for the
quantum noise of wave field is constant~\cite{loudon00}, that is,
position or momentum independence. Therefore, the measurement method
we have developed for coherent light is also applicable to coherent
state, making them a useful "testing ground" for quantum imaging and
mapping for information processing and quantum communication.
Recently, EPR entanglement in spatial coordinate has been
demonstrated by mixing an optical coherent light beam with squeezed
light~\cite{pklam08}. However, their measurements had involved
displacement and 'tilt' (momentum or angular) of a whole optical
beam, not the transverse amplitude and phase structure of the
optical beam. There is no approach in continuous variable quantum
mechanics to measure the Wigner function or density matrix of
spatial properties of EPR entangled beams. The optical technique and
procedure presented in this paper could provide detail studies of
phase-space physics in quantum metrology through sub-Planck
phase-space structures in the Wigner function, and also discrete
Wigner function for quantum mapping.

Phase-space distributions are used to represent quantum-mechanical
operators in exploring phase-space quantum effects and
quantum-classical correspondence. Quantum algorithms for measuring
KR and Wigner distributions have been
developed~\cite{saraceno04,saraceno02}. Discrete phase space
distribution has been suggested to show potential advantages in
quantum computing especially quantum mapping. Quantum algorithm can
be simply thought of as a quantum map acting in a Hilbert space of
finite dimensionality. Specifically, algorithms become interesting
in the large N limit ~i.e., when operating on many qubits!. For a
quantum map, this is the semiclassical limit where regularities may
arise in connection with its classical behavior. These semiclassical
properties may provide hints to develop new algorithms and ideas for
novel physics simulations.

In this paper, we demonstrate direct measurement of the KR
distribution for a wave field with Gaussian mode function of
$TEM_{oo}$. Linear transformation from KR distribution to Wigner, P
and Q-distribution are also plotted to show the fundamental
differences in theirs respective. Second, a superposition of two
spatially separated coherent light beams is used for discussing the
phase-space interferences in these distributions. We show that
complex conjugated KR distributions for spatial properties of wave
fields can be  determined by use of  a novel two-LO balanced
heterodyne detection scheme~\cite{kim99}. The technique measures
$\mathcal{K}^{*}(x,p)=<\mathcal{E}^{*}(x)\mathcal{E}(p)>e^{ixp}$,
which can be written as $S_{R}+iS_{I}$. A lock-in-amplifier is used
to measure the $S_{R}$ and $S_{I}$ with respect to relative phase
setting of a reference signal at $0^{\circ}$ and $90^{\circ}$,
respectively. By changing the lock-in-amplifier reference phases
such as $0^{\circ}$ and $-90^{\circ}$, the system will measure
$\mathcal{K}(x,p)$. However, we keep the lock-in-amplifier with
reference phases at $0^{\circ}$ and $90^{\circ}$ for all
measurements in this paper. There is no different in physics implied
by wave field in KR or complex conjugated KR representation. The
two-LO heterodyne technique was originally designed for biomedical
imaging, that is for optical phase space coherence tomography of the
light transmitted through or reflected from biological tissue. Now,
we use this measurement to explore the properties of one particle
wave mechanics or wave field through KR, Wigner, P- and Q-
distributions. The technique can be used to measure any complex
spatial wave fronts such as divergence and convergence, and
phase-conjugated properties of wave field~\cite{wax01,Reil05}. The
$\mathcal{K}^{*}(x,p)$ can be easily transformed to Wigner function
by using a linear transformation where the Radon transform is not
required. The advantage of KR is it contains local information of
the wave field. If there is no wave field presents at a
configuration space (x,p), then there will be no distribution at the
(x,p). It serves better for optical imaging in biomedical
application such as to characterize the cell structure. We will
illustrate this spatial property of KR distribution and compare it
with Wigner function, P- and Q- distributions. In the two-LO
balanced heterodyne detection, we use a local oscillator (LO) field
comprising a coherent superposition of a tightly focused LO Gaussian
beam of $TEM_{oo}$ and a highly collimated LO Gaussian beam of
$TEM_{oo}$.  This scheme permits independent control of the $x$ and
$p$ resolution, permitting concurrent localization of $x$ and $p$
with a variance product that surpasses the minimum uncertainty limit
associated with Fourier-transform pairs. Quantum mechanics did not
allow simultaneously measurement of x and p of a wave field.
However, simultaneous measurement in distribution of x and p for a
wave field is allowed.

\section{Characteristic Function Approach}

In this section, the characteristic function method will be used to
transform the characteristic function of KR distribution to Wigner,
P- and Q- distributions. The characteristic functions in term of
harmonic oscillator basis will be used through out this paper
because it is more revelent to spatial properties of coherent wave
fields. Since we directly measure the complex conjugated of KR
distribution, we will discuss how the measured results can be used
to obtain Wigner, P- and Q- distributions. We start with the
characteristic function for the complex conjugated of the KR
distribution in term of coherent state representation, as given by,
\begin{equation}
\mathcal{M}_{KR}(\beta,\beta^{*})= Tr(\rho
e^{-|\beta^{2}-\beta^{*2}|/4} e^{i\beta
\hat{a}^{\dag}+i\beta^{*}\hat{a}}). \label{eq:1}
\end{equation}
where ($\beta$, $\beta^{*}$) are the Fourier conjugate pairs for
($\alpha$, $\alpha^{*}$)the eigenvalues of $\hat{a}$ and
$\hat{a}^{*}$, respectively. The interesting feature of this
characteristic function in the Fourier plane $M_{KR}(\beta,
\beta^{*})$ is the trace of displacement operator
$\mathcal{\hat{D}}(\beta)$ followed by squeezing operator
$\mathcal{S}(1/2)$. The absolute of $|\beta^{2}-\beta^{*2}|/4$ just
to avoid the confusion of other forms of definition that are,
$\beta\rightarrow i\beta$ and $\beta^{*}\rightarrow -i\beta^{*}$.
The complex conjugated of KR distribution in the $(\alpha,
\alpha^{*})$ plane can be obtained through,
\begin{equation}
\mathcal{K}^{*}(\alpha, \alpha^{*})= \frac{1}{\pi^{2}}\int
d^{2}\beta e^{-i\beta
\alpha^{*}-i\beta^{*}\alpha}\mathcal{M}_{KR}(\beta,\beta^{*}).
\label{eq:2}
\end{equation}
The $\mathcal{K}^{*}(\alpha, \alpha^{*})$ can be transformed to
Wigner, P- and Q- distributions in term of $(\alpha,\alpha^{*})$
through the relationships of the characteristic functions, such
that,
\begin{equation}
\mathcal{M}_{KR}(\beta,\beta^{*})=e^{-|\beta^{2}-\beta^{*2}|/4}\mathcal{M}_{W}(\beta,\beta^{*})
\label{eq:3}
\end{equation}
where,
\begin{equation}
\mathcal{M}_{W}(\beta,\beta^{*})=e^{-|\beta|^{2}/2}\mathcal{M}_{P}(\beta,\beta^{*})=e^{|\beta|^{2}/2}\mathcal{M}_{Q}.
\label{eq:4}
\end{equation}
The characteristic function for Wigner function is given by,
\begin{equation}
\mathcal{M}_{W}(\beta,\beta^{*})= Tr(\rho e^{i\beta
\hat{a}^{\dag}+i\beta^{*}\hat{a}}). \label{eq:5}
\end{equation}
The P- and Q- representations are related to the characteristic
function of Wigner function through normal ordering and anti-normal
ordering of $(\hat{a},\hat{a}^{*})$.

In our experiment, we measure spatial properties of a wave field in
term of position and momentum coordinates (x,p). The
$\mathcal{K}^{*}(x,p)$ can be obtained from Eq.~\eqref{eq:1} and
Eq.~\eqref{eq:2} using the variables,
\begin{eqnarray}
\hat{a}=\frac{1}{\sqrt{2}}(\hat{x}+i\hat{p});\hat{a}^{\dag}=\frac{1}{\sqrt{2}}(\hat{x}-i\hat{p})\nonumber\\
\beta=\frac{1}{\sqrt{2}}(\sigma+i\tau);\beta^{*}=\frac{1}{\sqrt{2}}(\sigma-i\tau)\nonumber\\
\sigma=\frac{1}{\sqrt{2}}(x+ip);
\sigma^{*}=\frac{1}{\sqrt{2}}(x-ip).\label{eq:6}
\end{eqnarray}
Then, the following terms,
\begin{eqnarray}
e^{i\beta\hat{a}^{\dag}+i\beta^{*}\hat{a}}=e^{i\sigma\hat{x}+i\tau\hat{p}};\nonumber\\
e^{-|\beta^2-\beta^{*2}|/4}=e^{-i\sigma\tau/2};\nonumber\\
 e^{-i\beta\sigma^{*}-i\beta^{*}\sigma}=e^{-i\sigma x-i\tau
 p},\label{eq:7}
\end{eqnarray}
are obtained. The $\mathcal{K}^{*}(x,p)$ can be written as,
\begin{equation}
\mathcal{K}^{*}(x,p)=\int d\sigma d\tau e^{-i\sigma x-i\tau p}
Tr(\rho e^{i\sigma \hat{x}+i\tau \hat{p}}e^{i\sigma
\tau}).\label{eq:8}
\end{equation}
By using the identity $\int dx |x\rangle \langle x|=\hat{1}$ and
$\hat{\rho}=|\psi\rangle \langle \psi|$, we obtain,
\begin{equation}
\mathcal{K}^{*}(x,p)= \frac{2}{\pi}\psi^{*}(x) \int e^{-i\tau
p}\psi(x+\tau) d\tau. \label{eq:9}
\end{equation}
By changing the variable $\xi=x+\tau$, $d\xi =d\tau$, we obtain,
\begin{equation}
\mathcal{K}^{*}(x,p)=\frac{2}{\pi}\psi^{*}(x)\psi(p) e^{i x
p}.\label{eq:10}
\end{equation}
which is the complex conjugated of KR distribution. In order to
write the characteristic functions of Wigner, P- and Q-distributions
in term of spatial properties of coherent light beam such as beam
waist $\sigma$, position and momentum coordinates (x,p), we use the
variables,
\begin{equation}
\beta=\frac{1}{\sqrt{2}}(p'\sigma-i x'/\sigma);
 \beta^{*}=\frac{1}{\sqrt{2}}(p'\sigma+i
 x'/\sigma),\nonumber\\
\end{equation}
so that the characteristic functions for KR, Wigner, P- and Q-
distributions are related to each other as given by,
\begin{equation}
\mathcal{M}_{KR}(x,p)=e^{-i x' p'/2}\mathcal{M}_{W}(x',p')
\label{eq:11}
\end{equation}
and,
\begin{equation}
\mathcal{M}_{W}(x,p)=e^{-\frac{1}{4}(\sigma^2 p'^{2}+
x'^{2}/\sigma^{2})}\mathcal{M}_{P}(x',p') = e^{\frac{1}{4}(\sigma^2
p'^{2}+ x'^{2}/\sigma^{2})}\mathcal{M}_{Q}(x',p').\label{eq:12}
\end{equation}
Since our experiment measures the $\mathcal{K}^{*}(x,p)$, its
characteristic function is obtained through transformation as given,
\begin{equation}
\mathcal{M}_{KR}(x'p')=\int dx dp \mathcal{K}^{*}(x,p) e^{i x p'+i p
x'}.\label{eq:13}
\end{equation}
Then, the Wigner function is obtained through,
\begin{equation}
W(x,p)=\int dx' dp' e^{ix'p'/2} \mathcal{M}_{KR}(x',p')e^{-ix p'-ip
x'}\label{eq:14}
\end{equation}
or in the simplifying form as in the Ref ~\cite{kim99}, i.e.
\begin{equation}
W(x,p)=\frac{1}{\pi}\int dx'
dp'e^{-2i(x'-x)(p'-p)}\mathcal{K}^{*}(x,p).\label{eq:15}
\end{equation}
Similarly, the P- and Q- distributions can be obtained through,
\begin{equation}
P(x,p)=\int e^{i x' p'/2}\mathcal{M}_{KR}(x',p')
e^{\frac{1}{4}(\sigma^{2}p'^{2}+x'^{2}/\sigma^{2})} e^{-i x p'-i p
x'} dx' dp',\label{eq:16}
\end{equation}
and,
\begin{equation}
Q(x,p)=\int e^{i x' p'/2}\mathcal{M}_{KR}(x',p')
e^{-\frac{1}{4}(\sigma^{2}p'^{2}+x'^{2}/\sigma^{2})} e^{-i x p'-i p
x'} dx' dp',\label{eq:17}
\end{equation}
respectively.

The similarity of KR and Wigner function is  when they are
integrated over momentum/position, the two functions will provide
the same result for the probability in position/momentum i.e. $\int
\mathcal{K}^{*}(x,p) dx = \int W(x,p) dx =|\psi(p)|^2$ and $\int
\mathcal{K}^{*}(x,p) dp = \int W(x,p) dp =|\psi(x)|^2$. It should be
noted that because of the uncertainty principle, neither function
has a physical meaning until it is integrated over either momentum
space or configuration space.

\section{Experiment Details}

\begin{figure}
\begin{center}\
\includegraphics[scale=0.5]{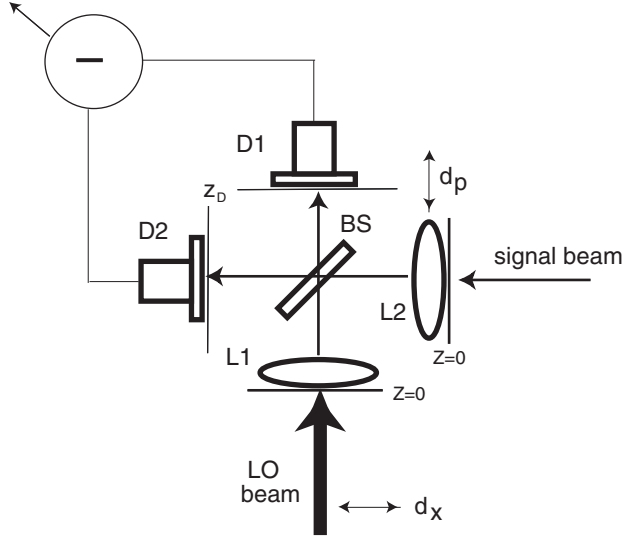}
\end{center}
\caption{A balanced heterodyne optical imaging scheme.}
\label{fig:briefsetup3}
\end{figure}

Since tomography methods developed for spatial properties of photon
wave function can be applied to classical field (coherent state with
large mean photon number), we use electric field notation
$\mathcal{E}(x)$ to represent the wave field $\psi(x)$ in the
following Sections.

We use a balanced heterodyne detection scheme as shown in
Fig.~\ref{fig:briefsetup3}. The beat amplitude $V_{B}$ is determined
by the  spatial overlap of the local oscillator $(LO)$ and  signal
$(S)$ fields in the plane of the detector at $Z = Z_{D}$ as,
\begin{equation}
V_{B}=\int
dx'\mathcal{E}^{*}_{LO}(x',z_{D})\mathcal{E}_{S}(x',z_{D})
\label{eq:18}
\end{equation}
where $x'$  denotes the transverse  position in the detector plane.
When the LO beam is moved by a distance  $d_{x}$, the $V_{B}$
becomes,
\begin{equation}
V_{B}(d_{x})=\int
dx'\mathcal{E}^{*}_{LO}(x'-d_{x},z_{D})\mathcal{E}_{S}(x',z_{D}).
\label{eq:19}
\end{equation}
The fields  in the detector plane are related to the  fields in the
source planes  $(z=0)$ of lenses L1 and L2, which have  equal focal
lengths  f = 6 cm.  The LO  and signal fields at $z=0$ after the
lenses, L1 and L2, are given by
\begin{eqnarray}
\mathcal{E}'_{LO}(x_{1}-d_{x},z=0)&=&\exp[-i\frac{k}{2f}x_{1}^{2}]\mathcal{E}_{LO}(x_{1}-d_{x},z=0)\nonumber\\
\mathcal{E}'_{S}(x_{2},z=0)&=&\exp[-i\frac{k}{2f}x_{2}^{2}]\mathcal{E}_{S}(x_{2},z=0)\nonumber\\
\label{eq:20}
\end{eqnarray}
When the lens L2 is scanned by a distance $d_{p}$, the signal
field~\eqref{eq:20} is altered as
\begin{equation}
\mathcal{E}'_{S}(x_{2},z=0)=\exp[-i\frac{k}{2f}(x_{2}-d_{p})^{2}]\mathcal{E}_{S}(x_{2},z=0)\nonumber\\
\end{equation}
The fields propagating a distance $d=f$ to the planes of  the
detectors can be obtained by using Fresnel's diffraction integrals
as,
\begin{eqnarray}
\mathcal{E}_{LO}(x'-d_{x},z_{D})&=&\sqrt{\frac{k}{i 2\pi f}}\int
dx_{1}\,\exp[i\frac{k}{2f}(x_{1}-x')^{2}]\nonumber\\
&\times&\exp[-i\frac{k}{2f}x_{1}^{2}]\,\mathcal{E}_{LO}(x_{1}-d_{x},z=0)\nonumber\\
\mathcal{E}_{S}(x',z_{D})&=&\sqrt{\frac{k}{i 2\pi f}}\int
dx_{2}\,\exp[i\frac{k}{2f}(x_{2}-x')^{2}]\nonumber\\
&\times&\exp[-i\frac{k}{2f}(x_{2}-d_{p})^{2}]\,\mathcal{E}_{S}(x_{2},z=0)\label{eq:21}
\end{eqnarray}

\begin{figure}
\begin{center}\
\includegraphics[scale=0.5]{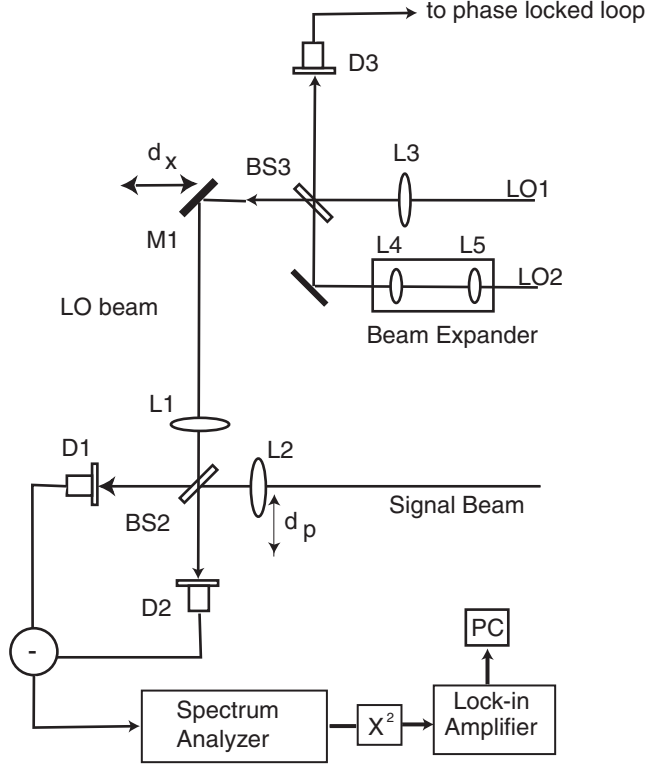}
\end{center}
\caption{Experimental setup for measuring the KR distribution using
two-LO balanced heterodyne detection. $X^{2}$; squarer, D;
photodiode detector, BS; beam splitter.} \label{fig:CH3TwoWindow}
\end{figure}

By substituting the above equations into Eq.~\eqref{eq:19}, the
quadratic phases in $x'$ cancel and the quadratic  phases that
depend on $x_{1,2}^{2}$ cancel in  these expressions because the
detector plane is in the focal plane of the lenses, L1 and L2. One
obtains
\begin{eqnarray}
V_{B}(d_{x},d_{p})&=&\frac{k}{f}\,\exp[i\frac{k}{2f}d_{p}^{2}]\int
dx_{2}\,\exp[-i\frac{k}{f}x_{2}d_{p}]\,\mathcal{E}_{S}(x_{2},z=0)\nonumber\\
&\times& \int
dx_{1}\,\mathcal{E}_{LO}^{*}(x_{1}-d_{x},z=0)\delta(x_{1}-x_{2})\label{eq:22}
\end{eqnarray}
Integrating over $x_{1}$ and by replacing $x_{2}$ by $x$ and
dropping the term $z=0$, the mean square beat amplitude is then
given by
\begin{equation}
|V_{B}(d_{x},d_{p})|^{2}\propto \left|\int dx\,
\mathcal{E}_{LO}^{*}(x-d_{x})\mathcal{E}_{S}(x)\,\exp(-ik\frac{d_{p}}{f}x)\right|^{2}\nonumber\\
\end{equation}
or
\begin{eqnarray}
|V_{B}(d_{x},d_{p})|^{2}&\propto& \int dx\,
\mathcal{E}_{LO}^{*}(x-d_{x})\mathcal{E}_{S}(x)\,\exp(-ik\frac{d_{p}}{f}x)\nonumber\\
&\times&\int dx'\,
\mathcal{E}_{LO}(x'-d_{x})\mathcal{E}_{S}^{*}(x')\,\exp(ik\frac{d_{p}}{f}x').\label{eq:23}
\end{eqnarray}
This can be rewritten using the variable transformations
$x=x_{o}+\eta/2$; $x'=x_{o}-\eta/2$ where the Jacobian  of this
transformation  is 1. Then, by using the definition of the Wigner
distribution,
\begin{equation}
W(x,p)=\int \frac{d\epsilon}{2\pi}\,\exp(i\epsilon p)\langle\,
\mathcal{E}^{*}(x+\epsilon/2)\mathcal{E}(x-\epsilon/2)\rangle
\label{eq:24}
\end{equation}
and its inverse transform is given by,
\begin{equation}
\mathcal{E}^{*}(x+\epsilon/2)\mathcal{E}(x-\epsilon/2)=\int dp\,
\exp(-i\epsilon p)W(x,p),
\end{equation}
the beat signal in Eq.~\eqref{eq:23} becomes
\begin{eqnarray}
|V_{B}(d_{x},d_{p})|^{2}&\propto& \int dx_{o} \int d\eta\,
\mathcal{E}_{LO}^{*}(x_{o}+\eta/2-d_{x})\mathcal{E}_{LO}(x_{o}-\eta/2-d_{x})\nonumber\\
&\times& \int dp\, \exp(-ik\frac{d_{p}}{f}\eta)\,\exp(-i\eta
p)W_{S}(x,p). \label{eq:25}
\end{eqnarray}
Since the Wigner distribution of the LO field is
\begin{eqnarray}
W_{LO}(x_{\circ}-d_{x},p+k\frac{d_{p}}{f})&=&\int
\frac{d\eta}{2\pi}\,
\exp[-i\eta(p+k\frac{d_{p}}{f})]\nonumber\\
&\times&\mathcal{E}_{LO}^{*}(x_{o}+\eta/2-d_{x})\mathcal{E}_{LO}
(x_{o}-\eta/2-d_{x}) \label{eq:26}
\end{eqnarray}
then the LO fields in Eq.~\eqref{eq:25} can be replaced by the
Wigner function in Eq.~\eqref{eq:26}. Finally, the mean square
heterodyne beat signal $|V_{B}|^{2}$ can now be written as
\begin{equation}
|V_{B}(d_{x},d_{p})|^{2} \propto \int dx dp \,W_{LO}(x-d_{x},
p+\frac{k}{d_{p}})W_{S}(x,p)\label{eq:27}
\end{equation}
where $W_{S}(x,p)$  is the Wigner distribution  of the signal field
in the plane of L2  (z =0) and $W_{LO}(x,p)$ is the LO Wigner
distribution in the  plane of L1 (z = 0). We include a detail
description of two-window heterodyne measurement of KR distribution
as shown in Fig.~\ref{fig:CH3TwoWindow}. The variables $d_{x}$ and
$d_{p}$ respectively indicate the positions of  a mirror M1 and a
lens L2 as in Fig.~\ref{fig:CH3TwoWindow}. Eq.~\eqref{eq:27} shows
that the mean-square beat  signal yields a phase-space contour plot
of $W_{S}(x,p)$ with phase space resolution determined  by
$W_{LO}(x,p)$. By using a two-LO heterodyne detection scheme as
discussed below, the $|V_{B}(d_{x},d_{p})|^{2}$ is found to be
proportional to $\mathcal{K}^{*}(x,p)$.

\begin{figure}
\begin{center}\
\includegraphics[scale=0.5]{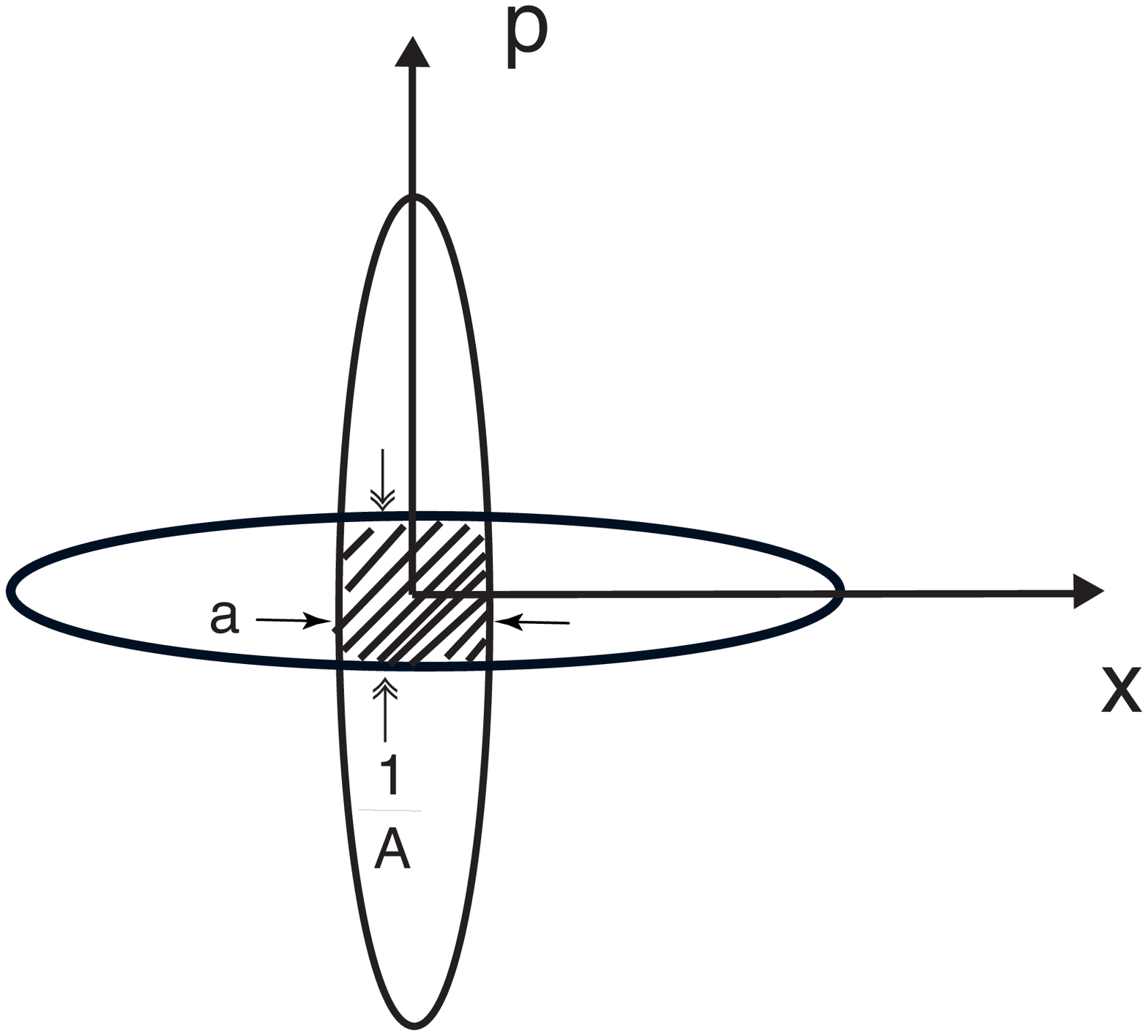}
\end{center}
\caption{The overlapping area is the position and momentum
resolutions of the combined LO beams. The uncertainty is
proportional to $\frac{a}{A}$ less than 1.} \label{fig:smallandbig}
\end{figure}
To obtain independent control of the $x$ and $p$ resolution in
heterodyne measurement, we employ a slowly varying LO field
containing a focused and a collimated field with a well defined
relative phase $\theta$
\begin{equation}
\mathcal{E}_{LO}(x)=\mathcal{E}_{\circ}[\,\exp(-\frac{x^{2}}{2a^{2}})+\alpha\,
\exp(-\frac{x^{2}}{2A^{2}})\,\exp(i\theta)] \label{eq:28}
\end{equation}
Here $a$  is chosen  to be  small compared with  the distance scales
of interest  and $1/A$ is  chosen to  be small  compared with the
momentum scales of  interest in the signal field. The schematic
picture of the overlapping of the LO beam with the spatial width $a$
and another LO beam with the spatial width $A$ is shown in
Fig.~\ref{fig:smallandbig}. One can see that the overlapping area is
determined by the position and momentum resolutions for the LO
fields in Eq.~\eqref{eq:28}. The focussed LO gaussian beam extracts
the position information of the signal field and the collimated LO
gaussian beam extracts momentum information of the signal field. The
Wigner function for the LO field is obtained by substituting
Eq.~\eqref{eq:28} into Eq.~\eqref{eq:24}. We take
$A^{2}$$\gg$$a^{2}$. In this case the phase-($\theta$-) dependent
part of the Wigner function for the LO takes the form
\begin{eqnarray}
W_{LO}(x,p)&\propto& exp(-\frac{2x^{2}}{A^{2}}-2a^{2}p^{2})\cos(2xp + \theta)\nonumber\\
           &\simeq&\cos(2xp + \theta),\label{eq:29}
\end{eqnarray}
where  the  last form  assumes  that the  range  of  the momentum
and position integration in relation~\eqref{eq:27} is limited by the
signal field.

The measurement of phase space distributions is accomplished by
translation of optical  elements. These elements are all mounted on
translation stages  driven by computer controlled linear actuators.
The system  scans the LO position over a distance  $d_{x}$ = $\pm 1$
cm by translating mirror M1 in the LO path. The LO  momentum is
scanned over $\pm 0.3\,k$, where $k = 2\pi/\lambda$ is an optical
wave vector, by translation of the signal-beam input lens L2 (focal
length $f$ = 6 cm) by a distance $d_{p}$.

In the experiments, as illustrated  in Fig.~\ref{fig:CH3TwoWindow},
the LO beam is obtained by combination  of two fields that differ in
frequency by 5  kHz, so that $\theta$ =$(2\pi \times 5kHz)t$. Lens
L3 focuses beam LO1 to a waist of width $a$, and lenses L4 and L5
expand beam LO2 to width $A$. We combine these two components at
beam splitter BS3 to obtain an LO field of the form given in
Eq.~\eqref{eq:28}. We monitor one output of the beam splitter with
detector  $D_{3}$ to phase lock  the 5 kHz beat signal to  the
reference channel of the lock-in amplifier. Each component of the LO
beam is shaped so that it is at a  beam waist at the input plane of
the heterodyne imaging system (lens L1). The focussed gaussian LO
beam is frequency shifted at 110 MHz and the collimated gaussian LO
beam is frequency shifted at 110 MHz plus 5 kHz. These two LO beams
are overlapped with each other and phase locked at 5 kHz. The signal
beam is frequency shifted at 120 MHz. Two imaging lenses L1 and L2
are used to overlap the dual LO beam with the signal beam at two
detectors. The interference beat signal between the signal beam and
the dual LO beam is obtained at detectors 1 and 2 and consists of 10
MHz and 10 MHz plus 5 kHz components. These signals are sent to a
spectrum analyzer. The spectrum analyzer bandwidth, 100 kHz, is
chosen to be large compared with 5 kHz difference frequency. The
output of the analyzer is then squared by using an analog
multiplier. The mean square signal has components at 5 kHz. A
lock-in amplifier is used to measure the in- and out-of phase
components of the multiplier output at 5 kHz. The lock-in outputs
for  in- and out-of-phase quadratures  then directly determine the
real and the imaginary parts of the $\mathcal{K}^{*}(x,p)$ function
\begin{eqnarray}
|V_{B}(d_x, d_p)|^{2}=\mathcal{K}^{*}(x_{\circ},p_{\circ})&=&\int \frac{dx'dp'}{\pi}\,\exp[2i(x'-x_{\circ})\times (p'-p_{\circ})]W_{S}(x',p')\nonumber\\
                    &=&\langle\mathcal{E}^{*}(x_{\circ})\mathcal{E}(p_{\circ})\rangle\,
                    \exp(ix_{\circ}p_{\circ})=S_{R}+i S_{I}, \label{eq:30}
\end{eqnarray}
where the $W_{LO}(x,p)$ in Eq.~\eqref{eq:29} is replaced by
$e^{i(2xp+\theta)}$ in Eq.~\eqref{eq:27} yielding the in-phase and
out-of-phase contributions from $\theta$. Here $x_{\circ}$ = $d_{x}$
is the center position of the LO  fields and $p_{\circ}$ =
$-kd_{p}/f$ is the  center momentum. The $S_{R}$ and $S_{I}$
components are related to heterodyne beat signal of
$E^{*}_{LO1}E_{S}$ and $E^{*}_{LO2}E_{S}$ or intensity correlation
of $E^{*}_{LO1}E_{S}$ and $E^{*}_{LO2}E_{S}$. In the balanced
heterodyne detection, the output voltage
$V_{B}(t)$=$V_{1}(t)-V_{2}(t)$ before being fed into the spectrum
analyzer is given by,

\begin{equation}
V_{B}(t)=2E_{S}^{*}E_{LO1}e^{-i(\Omega_{LO1}-\Omega_{S})t}+2E_{S}^{*}E_{LO2}e^{-i(\Omega_{LO2}-\Omega_{S})t}+cc\label{eq:31}
\end{equation}
In spectrum analyzer, the power spectrum of the $V_{B}$ is measured
as

\begin{equation}
|V_{B}(\Omega)|^{2}=\int_{-\infty}^{\infty}\frac{d\tau}{2\pi}\,e^{i\Omega\tau}\,\langle
V_{B}(t)V_{B}(t+\tau)\rangle .\label{eq:32}
\end{equation}
As mentioned previously, it is squared by using a squarer to recover
the beat signal $|V_{B}(\Omega)|^2 $. From Eq.~\eqref{eq:32}, the
power spectrum can be calculated by keeping the slowly varying term
$(\Omega_{LO1}-\Omega_{LO2})$ in time $t$ and other terms that
depend on $\tau$, that is,

\begin{eqnarray}
\langle V_{B}(t)V_{B}(t+\tau)\rangle&=&\frac{1}{T}\int_{-T/2}^{T/2}dt\,V_{B}(t)V_{B}(t+\tau)\nonumber\\
          &\propto&E_{S}^{*}E_{LO1}E_{S}E_{LO1}^{*}e^{-i(\Omega_{S}-\Omega_{LO1})\tau}+E_{S}^{*}E_{LO2}E_{S}E_{LO2}^{*}e^{-i(\Omega_{S}-\Omega_{LO2})\tau}\nonumber\\
          &+&E_{S}^{*}E_{LO1}E_{S}E_{LO2}^{*}e^{-i(\Omega_{LO1}-\Omega_{LO2})t}e^{-i(\Omega_{S}-\Omega_{LO2})\tau}\nonumber\\
          &+&E_{S}^{*}E_{LO2}E_{S}E_{LO1}^{*}e^{i(\Omega_{LO1}-\Omega_{LO2})t}e^{-i(\Omega_{S}-\Omega_{LO1})\tau}\nonumber\\
          &+&(\textrm{: the negative frequency contributions from the above terms:})\nonumber\\
\label{eq:33}
\end{eqnarray}
Here, $\Omega_{S}$=120 MHz, $\Omega_{LO1}$=110 MHz + 5 kHz and
$\Omega_{LO2}$=110 MHz. Now, by substituting Eq.~\eqref{eq:33} into
Eq.~\eqref{eq:32} to obtain the power spectrum for the beat $V_{B}$
and setting the analyzer at 10 MHz with the bandwidth of 100 kHz,
the $|V_{\circ}(t)|^2$ at 5 kHz after the recovery by the squarer is

\begin{equation}
|V_{\circ}(t)|^{2}\propto
E_{S}^{*}E_{LO1}E_{S}E_{LO2}^{*}e^{-i(\Omega_{LO1}-\Omega_{LO2})t}+cc
\label{eq:34}
\end{equation}
Here $\Omega_{LO1}-\Omega_{LO2}$ = 5 kHz. The in-phase and
out-of-phase components of the $|V_{\circ}(t)|^{2}$ at 5 kHz
correspond to $S_{R}$ and $S_{I}$ in Eq.~\eqref{eq:30}. Note that
$E_{S}^{*}E_{LO1}$ is integrated over the transverse plane as is
$E_{S}E^{*}_{LO2}$. It is worth noting that the component
$E_{S}^{*}E_{LO1}$ of Eq.~\eqref{eq:34} is corresponding to the
measurement of the position distribution in
$K^{*}(x_{\circ},p_{\circ})$ of Eq.~\eqref{eq:30} by the tightly
focussed LO1 beam. Similarly, the component $E_{S}E_{LO2}^{*}$ of
Eq.~\eqref{eq:34} indicates the measurement of the momentum
distribution in $K^{*}(x_{\circ},p_{\circ})$ of Eq.~\eqref{eq:30} by
the collimated LO2 beam.

As the  position of mirror  M1 is scanned a distance $d_{x}$, the
optical path lengths of the LO fields change. For the current
experiments, the HeNe laser is a source, the change in path lengths
is small compared with the  Rayleigh  length  and  the coherence
length  of  the  beams,  so translating  M1   simply changes  the
center  position   of  the  LO fields.

\section{Results}

\subsection{Measurement of an Optical Gaussian beam}

\subsubsection{Gaussian Beam}

\begin{figure}
\begin{center}\
\includegraphics[scale=0.5]{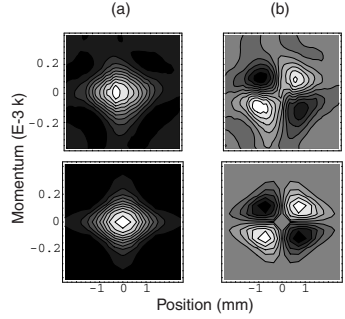}
\end{center}
\caption{The Wigner function for a gaussian beam. Top row is
experimental results and bottom row is theoretical prediction for a
gaussian beam. (a) is in-phase component of
$\mathcal{K}^{*}(x_{\circ},p_{\circ})$ and (b) is out-of-phase
component of $\mathcal{K}^{*}(x_{\circ},p_{\circ})$}.
\label{fig:Gaussian}
\end{figure}

As  an initial  demonstration of  the  capability of  this system,
we measure the $\mathcal{K}^{*}(x,p)$ function for  an ordinary
gaussian beam. A one dimensional wave field for an Gaussian beam of
$TEM_{00}$ and radii of curvature, R, is given by,
\begin{eqnarray}
\mathcal{E}(x) &\propto&
\exp({-\frac{x^2}{2\sigma^{2}}+i\frac{kx^{2}}{2R}})\nonumber\\
&\propto&\mathcal{A} + i \mathcal{B}\label{eq:35}
\end{eqnarray}
where the x is the transverse position and $\sigma$ is the waist of
signal beam. The Fourier transform of $\mathcal{E}(x)$ is,
\begin{eqnarray}
\mathcal{E}(p)&\propto&
\exp({-\frac{p^2}{8\sigma^{2}(\frac{1}{2\sigma^2})^2+(\frac{k}{2R})^2}}
{-i\frac{k p^2}{8 R
(\frac{1}{2\sigma^2})^2+(\frac{k}{2R})^2}})\nonumber\\
&\propto&\mathcal{C}+i\mathcal{D}\label{eq:36}
\end{eqnarray}
The complex conjugated KR distribution for this wave field can be
written as,
\begin{eqnarray}
\mathcal{K}^{*}(x,p)&=&(\mathcal{AC+BD}) \cos(xp)+(\mathcal{BC-AD})\sin(xp)\nonumber\\
&+& i((\mathcal{AD-BC}) \cos(xp)+(\mathcal{AC-BD})
sin(xp))\label{eq:37}
\end{eqnarray}
The characteristic function, $\mathcal{M_{KR}}(x,p)$, is obtained
from Eq.~\eqref{eq:11}, where the characteristic function for Wigner
function for this wave field is given by,
\begin{eqnarray}
\mathcal{M_{W}}(x',p')&=&\int \mathcal{E}^{*}(\eta
-\frac{x'}{2})\mathcal{E}(\eta+\frac{x'}{2})e^{ip'\eta}
d\eta \nonumber\\
 &=&\int \mathcal{E}^{*}(\eta
+\frac{p'}{2})\mathcal{E}(\eta-\frac{p'}{2})e^{ix'\eta}
d\eta \nonumber\\
&=&\exp(-\frac{x'}{4\sigma^{2}}-\frac{\sigma^2}{4}(\frac{kx'}{R}+p')^2)\label{eq:38}
\end{eqnarray}
Then, the Wigner, P- and Q-distributions are obtained from
Eq.~\eqref{eq:14}, Eq.~\eqref{eq:16} and Eq.~\eqref{eq:17},
respectively. The Wigner function for the wave field is given by,
\begin{eqnarray}
W(x,p)=\frac{1}{\pi}\exp(-\frac{x^2}{\sigma^{2}}-\sigma^2(\frac{kx}{R}+p)^2)\label{eq:39}
\end{eqnarray}
The P- and Q-distributions for the wave field are given in the
integral form as,
\begin{eqnarray}
P(x,p)&=&\int \exp(-\frac{\sigma^2 k^2 x'^2}{4
R^2}-\frac{\sigma^2 k x' p'}{2R})exp(-ixp'-ipx') dx'dp' \nonumber\\
Q(x,p)&=&\int \exp(-\frac{x'^2}{2\sigma^2}-\frac{\sigma^2
p'^2}{2}-\frac{\sigma^2 k^2 x'^2}{4 R^2}-\frac{\sigma^2 k x'
p'}{2R})\exp(-ixp'-ipx') dx' dp'\nonumber\\
\label{eq:40}
\end{eqnarray}
For simplicity, the signal Gaussian beam is shaped  by a telescope
so that its waist coincides with input plane L2 of the heterodyne
imaging system. For a gaussian beam at its waist, $R =\infty $,
Eq.~\eqref{eq:37} gives the complex conjugated KR distribution as,
\begin{equation}
\mathcal{K}^{*}(x,p)=\mathcal{AC}\cos(xp)+i\mathcal{AC}\sin(xp)
 \label{eq:41}
\end{equation}
where $\mathcal{A}=\exp(-\frac{x^2}{2\sigma^2})$,
$\mathcal{C}=\exp(-\frac{p^2\sigma^2}{2})$, and  $\sigma$=0.85 mm is
the $1/e$-intensity width. The $\mathcal{K}^{*}(x,p)$ function for
the signal field is measured by use of the dual LO beam of the form
given by Eq.~\eqref{eq:28} with $a$ =81 $\mu$m, $A$ = 2.6 mm, and
$\alpha$ =1.  The measurement result for a gaussian beam is shown in
Fig.~\ref{fig:Gaussian}. The top row is our experimental results and
the bottom row is a theoretical prediction obtained by using
Eq.~\eqref{eq:41}. The real and the imaginary parts of the detected
signal, Eq.~\eqref{eq:30}, are shown in Fig.~\ref{fig:Gaussian}(a)
and (b). Our observation is similar to the theoretical prediction by
Wodkiewicz~\cite{wodkiewicz03b} for a coherent state.

\begin{figure}
\begin{center}\
\includegraphics[scale=0.5]{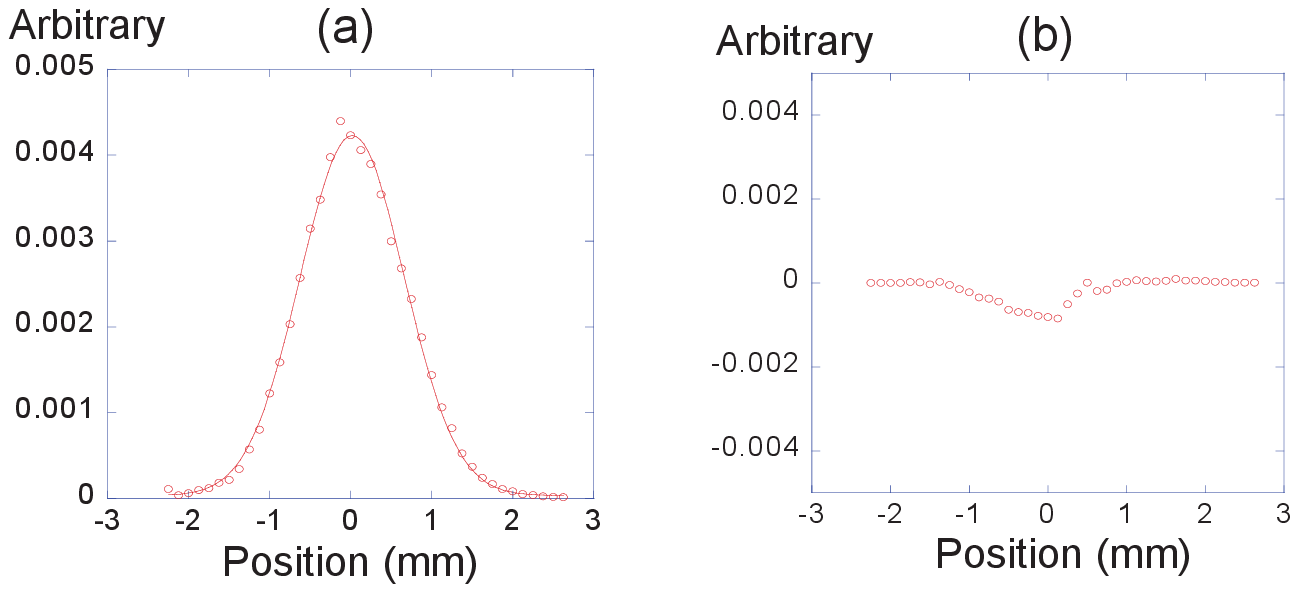}
\end{center}
\caption{(a) Real part and (b) imaginary part of position
distribution of a Gaussian field. Imaginary part is around zero.}
\label{fig:KRpositionGauss}
\end{figure}

\begin{figure}
\begin{center}\
\includegraphics[scale=0.5]{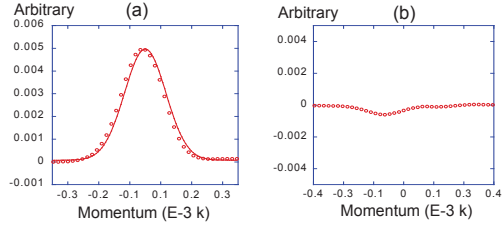}
\end{center}
\caption{a) Real part and (b) imaginary part of momentum
distribution of a Gaussian field. Imaginary part is around zero.}
\label{fig:KRMomentumGauss}
\end{figure}

Position (momentum) distributions of this field can be obtained
through the summation of momentum (position) coordinate of real and
the imaginary parts of the measured $\mathcal{K}^{*}(x,p)$.  The
position and momentum distributions are shown in
Fig.~\ref{fig:KRpositionGauss} and Fig.~\ref{fig:KRMomentumGauss},
respectively. The imaginary part of position and momentum
distribution are around zero as theoretically predicted by $\int
\mathcal{K}^{*}(x,p) dp=|\mathcal{E}(x)|^2$ and $\int
\mathcal{K}^{*}(x,p) dx=|\mathcal{E}(p)|^2$, respectively, which are
the real physical quantities (no complex values). From here, the
position and momentum distributions are fitted with Gaussian
function. We obtain the beam waist of $\sigma$ = 0.86 from position
distribution and $\sigma$=0.87 from momentum distribution. Both
results are in excellent agreement with the measured width
$\sigma$=0.85 mm obtained by use of a diode array, demonstrating
that high position and momentum resolution can be jointly obtained.

\begin{figure}
\begin{center}\
\includegraphics[scale=0.5]{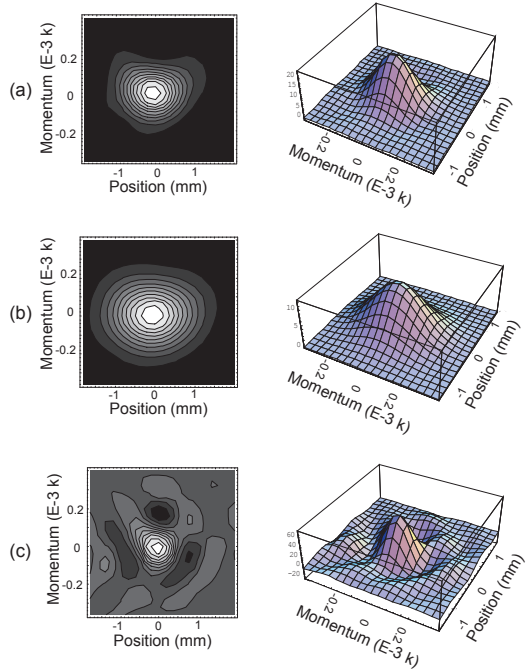}
\end{center}
\caption{(a) The reconstructed Wigner distribution, (b) Husimi or
Q-distribution, and (c) Glauber Sudarshan P-distribution for an
optical Gaussian beam. 2D-plot(left);3D-plot(right).}
\label{fig:PWQGaussian}
\end{figure}

Wigner distribution is obtained by using simple linear
transformation as discussion in Eq.~\eqref{eq:14} or
Eq.~\eqref{eq:15} as shown in Fig.~\ref{fig:PWQGaussian}(a). We
$\chi^{2}$ fit the width of the measured in-phase signal  in
position for $p$ = 0 to obtain a spatial  width  $\sigma$  =0.87
mm, whereas the corresponding momentum distribution  for  $x$ =  0
yields $\sigma$ =0.83 mm.

The characteristic function $\mathcal{M}_{KR}(x',p')$ is obtained by
using numerical integration of the measured $\mathcal{K}^{*}(x.p)$
as in Eq.~\eqref{eq:13}. The P- and Q- distributions for this signal
gaussian beam are then obtained through the Eq.~\eqref{eq:16} and
Eq.~\eqref{eq:17}, as shown in Fig.~\ref{fig:PWQGaussian}(b) and
Fig.~\ref{fig:PWQGaussian} (c), respectively. The P-distribution has
a narrower peak in phase-space compared to other distributions. This
is predicted for the signal beam with $R =\infty $ in P-distribution
of Eq.~\eqref{eq:40}, which we should have $\delta(x) \delta(p)$ in
phase-space. The Q-distribution has a broad peak in phase-space
compared to other distributions. The Q-distribution for this signal
beam can be evaluated from Eq.~\eqref{eq:40} with $R=\infty$, as
given by,
\begin{equation}
Q(x,p)\propto \exp(-\frac{p^2 \sigma^2}{2}-\frac{x^2}{2\sigma^2}).
 \label{eq:42}
\end{equation}
The position width of the Q-distribution at p=0 is about $\sqrt{2}$
larger than the position width of Wigner distribution at p=0 as from
Eq.~\eqref{eq:39}. Hence, the beam waist for the signal beam
obtained from the Q-distribution is $\sqrt{2}$ larger than the exact
value.

\subsubsection{Measurement of superposition of two slightly displaced coherent beams}

\begin{figure}
\begin{center}\
\includegraphics[scale=0.5]{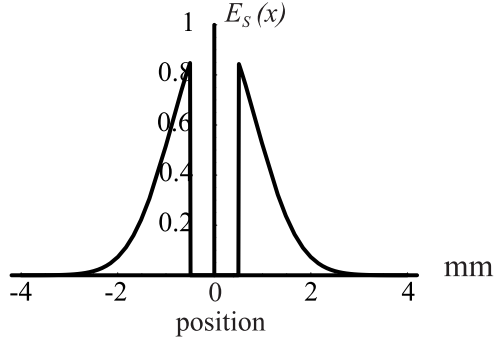}
\end{center}
\caption{The wave field $\mathcal{E}(x)$ of a Gaussian beam obscured
by a wire.} \label{fig:CH3WireFunction}
\end{figure}

\begin{figure}
\begin{center}\
\includegraphics[scale=0.5]{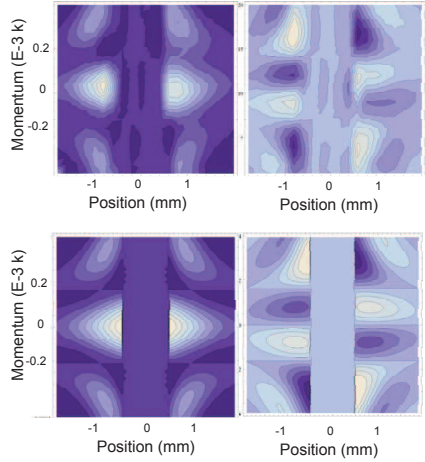}
\end{center}
\caption{Top row shows experimental results and bottom row shows
theoretical predictions for a gaussian beam blocked by a wire. (a)
in-phase component of the $\mathcal{K}^{*}(x_{\circ},p_{\circ})$ and
(b) out-of-phase component of the
$\mathcal{K}^{*}(x_{\circ},p_{\circ})$.} \label{fig:CH3WireKR}
\end{figure}

As a fundamental feature in the process of quantum measurement, we
cannot observe physical properties of a quantum objects directly
because the overall backaction of any observation cannot be made
less than Planck's constant h. Instead, we observe the wave or the
particle aspects of the physical objects. The $\mathcal{K}^{*}(x,p)$
distribution for a coherent field is more likely representing
particle picture of the field because it contains local information
of the coherent field. While the Wigner function is more likely
representing wave behavior of the field because it exhibits
phase-space interference. Based on these properties, the
$\mathcal{K}^{*}(x,p)$ and Wigner distributions are very useful to
characterize a wave field or an physical object through phase-space
imaging in many applications such as quantum imaging, metrology, and
biomedical imaging. To illustrate the particle picture of
$\mathcal{K}^{*}(x,p)$ and the wave picture of Wigner function, we
use the same signal gaussian beam obscured by a wire with diameter
of 1 mm. Then, the electric field $\mathcal{E}_{s}(x)$ as a function
of position is shown in Fig.~\ref{fig:CH3WireFunction}. It didn't
involve convolution integration because the wire is placed close to
the imaging lens L2. It shows the superposition of two slightly
displaced coherent beams. It is analogous to a Schr\"{o}dinger cat
state. In this case, the slowly varying field is gaussian  as before
but multiplied by a slit function that sets the field equal to zero
for $|x|\leq 0.5$ mm. Fig.~\ref{fig:CH3WireKR} shows contour plots
of the real and the imaginary parts of  the detected signal of
$\mathcal{K}^{*}(x,p)$.

\begin{figure}
\begin{center}\
\includegraphics[scale=0.5]{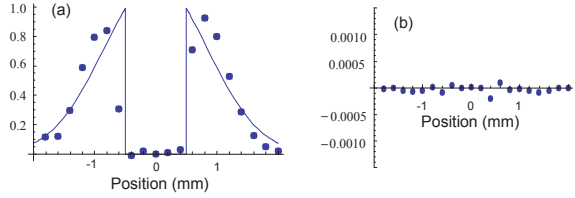}
\end{center}
\caption{(a) The real part, (b) the imaginary part of the position
distribution of a wire function obtained by integrating the measured
KR distribution over momentum. Imaginary part is around zero.}
\label{fig:KRpositionWire}
\end{figure}

\begin{figure}
\begin{center}\
\includegraphics[scale=0.5]{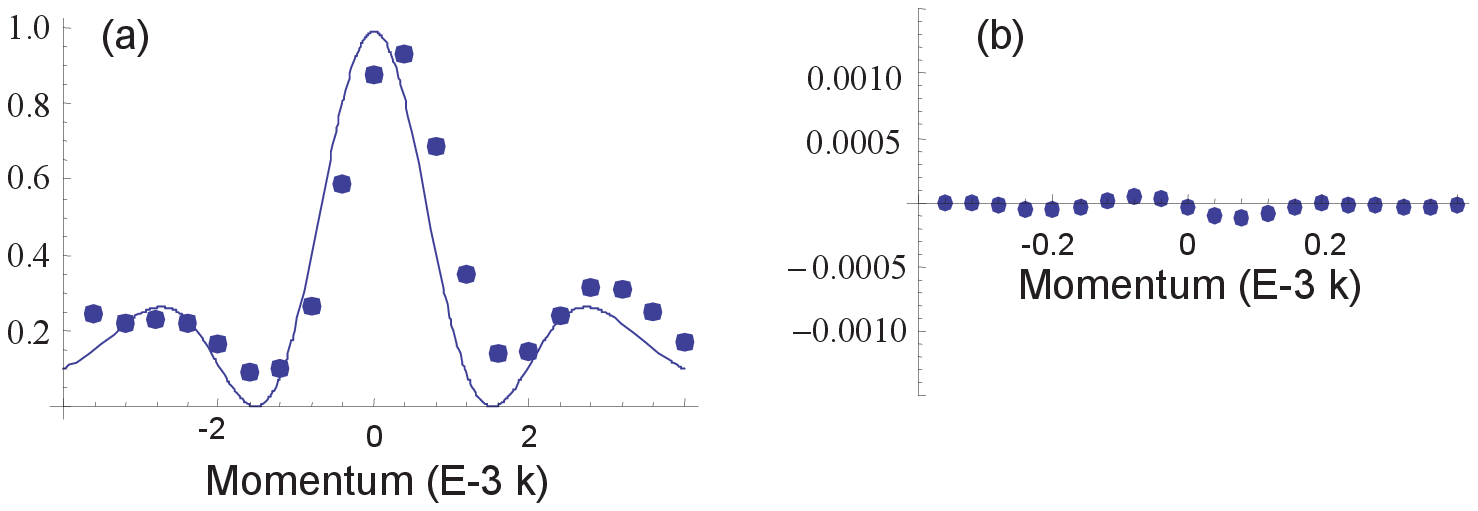}
\end{center}
\caption{(a) The real part, (b) the imaginary part of the momentum
distribution of a wire function obtained by integrating the measured
KR distribution over position. Imaginary part is around zero.}
\label{fig:KRMomentumWire}
\end{figure}

The top row is our experimental results. The bottom row is
theoretical prediction. The theoretical plots are first obtained by
numerically generating the signal function $\mathcal{E}_{S}(x)$ and
its Fourier transform $\mathcal{E}_{S}(p)$. The real and imaginary
parts of $\mathcal{K}^{*}(x,p)$ in Eq.~\eqref{eq:10} are then
theoretically plotted. The real and imaginary parts are not showing
phase-space interferences of two slightly displaced coherent beams.
These plots show local information of the signal field, i.e.,
zero(nonzero) field is corresponding to zero(nonzero)
$\mathcal{K}^{*}(x,p)$ phase-space distribution. This locality
property exhibits particle picture if an atomic wave function/single
photon function is used. The position and momentum distributions are
then retrieved $\int\mathcal{K}^{*}(x,p)dp$ and
$\int\mathcal{K}^{*}(x,p)dx$ as shown in
Fig.~\ref{fig:KRpositionWire} and Fig.~\ref{fig:KRMomentumWire},
respectively. The momentum distribution contains the interference
features of two spatially separated wave packets of
$\mathcal{E}_{S}(x)$ as shown in Fig.~\ref{fig:KRMomentumWire}. The
imaginary part of these distributions are around zero as expected.

\begin{figure}
\begin{center}\
\includegraphics[scale=0.5]{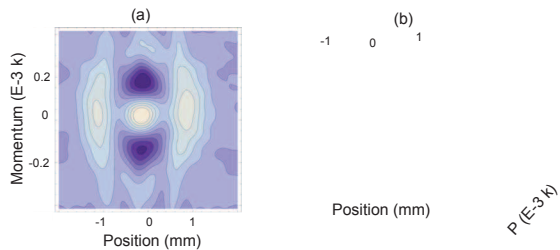}
\end{center}
\caption{The reconstructed Wigner distribution for a wire function
(a) two-, (c) three- dimensional plot.} \label{fig:WignerWire}
\end{figure}

The Wigner function as shown in Fig.~\ref{fig:WignerWire} is
reconstructed by using the linear transformation of the measured
$\mathcal{K}^{*}(x,p)$. The coherence between these two wave packets
in the signal field leads to an phase-space interference pattern in
the momentum distribution. The signature of this coherence in the
Wigner distribution is the oscillating positive and negative values
between the main lobes. An interesting feature of this Wigner
distribution is the oscillation in momentum at the position $x$ =0
of the wire. We observe the negative values which is analogous to
quantum interference in phase space. This feature can be seen in
Fig.~\ref{fig:WignerWire},  in which the reconstructed Wigner
function is shown as a three dimensional plot. The negative values
highlight the impossibility of a particle simultaneously having a
precise position and momentum. It also makes sure that the sum over
momentum along x = 0 in the reconstructed Wigner distribution has
zero intensity at the center. The negative and positive parts of the
Wigner phase space distribution are important features to obtain
full information about the field. Our observation of the negative
values of Wigner function for this coherent field did not claim that
the negative values exhibited by a quantum field is not a quantum
feature. We believe that classical or quantum features of an
experiment are based on classical or quantum field involved in the
experiment.

\begin{figure}
\begin{center}\
\includegraphics[scale=0.5]{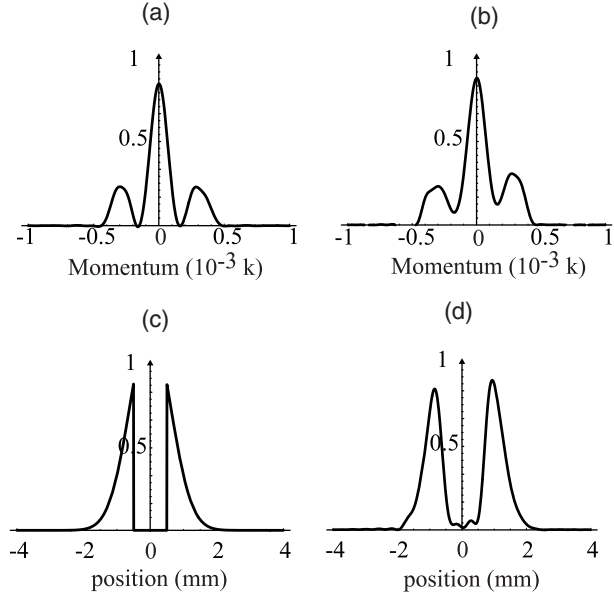}
\end{center}
\caption{The momentum and position distribution for the
Schr\"{o}dinger cat state. (a) and (c) are theoretical prediction of
momentum and position distributions of the cat state. (b) and (d)
are the corresponding experimental results of momentum and position
distribution of the cat state obtained by integrating the measured
Wigner distribution over $p$ and $x$ respectively.}
\label{fig:XPDist}
\end{figure}

From the obtained Wigner function, one can obtain the position and
momentum distribution of the $\mathcal{E}_{s}(x)$ by using the
formulas
$|\mathcal{E}_{S}(p)|^{2}$=$\int_{-\infty}^{\infty}W_{S}(x,p)\,dx$
and
$|\mathcal{E}_{S}(x)|^{2}$=$\int_{-\infty}^{\infty}W_{S}(x,p)\,dp$
 as shown in Fig.~\ref{fig:XPDist}(b) and (d) respectively.
The resolution of these plots is better than
Fig.~\ref{fig:KRpositionWire} and Fig.~\ref{fig:KRMomentumWire}
because we numerically generate more data points from the
reconstructed Wigner function.
 The measurements are in agreement with the theoretical
predictions as shown in Fig.~\ref{fig:XPDist}(a) and (c)
respectively.

\begin{figure}
\begin{center}\
\includegraphics[scale=0.5]{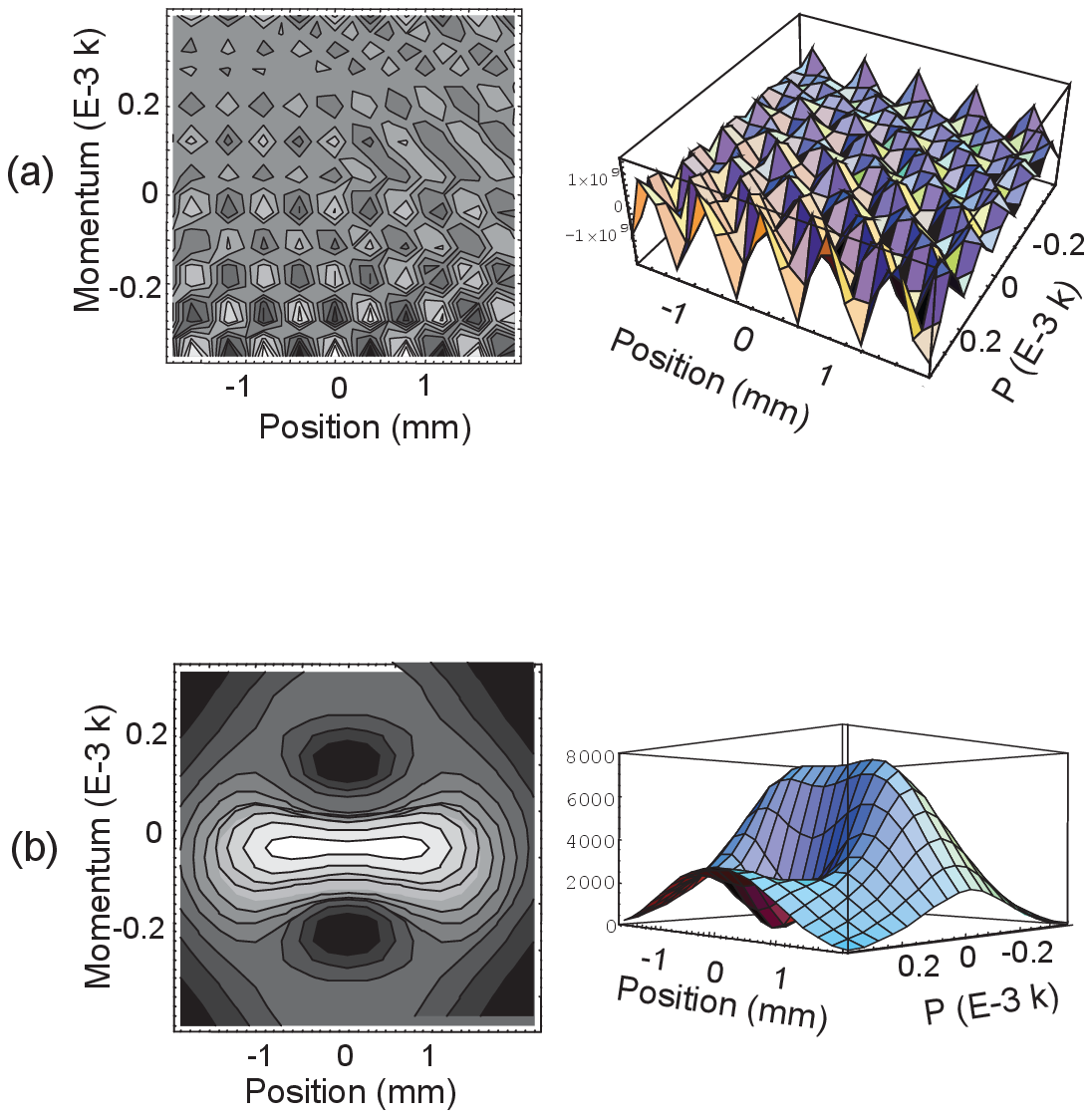}
\end{center}
\caption{The reconstructed (a) P- and (b) Q-distributions for a
slightly displaced coherent beams. 2D-plot(left); 3D-plot(right).}
\label{fig:PQwireGauss}
\end{figure}

As discussed before, we first obtain the characteristic function
$\mathcal{M}_{KR}(x',p')$ and then the P- and Q- distributions for
this signal field. These distributions are plotted in
Fig.~\ref{fig:PQwireGauss} (a) and (b), respectively. The
Q-distribution exhibits the broadening or low resolution of
phase-space features compared to Wigner function. While the
P-distribution is not a well behaved function for this field as it
is expected for P-distribution.

\section{Discussion and Conclusions}

Spatial properties of photon wave mechanics approach~\cite{smith07}
for a single photon have been studied in detail. The approach can be
applied to both single photon state and coherent field. As a
consequence, the coherent field is the best testing ground for
developing tomography method for quantum information processing. In
the similar efforts, coherent fields have played an important role
in quantum communication and computing such as search
algorithm~\cite{lloyd00,spreeuw02,spreeuw01,kim09} and factorization
of numbers~\cite{schleich08}. Optical wave mechanics
implementations~\cite{kim04,kim02} of entanglement and superposition
with coherent fields (coherent state with large photon numbers) have
been demonstrated. This implementation has been used to study
entanglement swapping and tests of non-locality. However, photon
wave mechanic approach for two- and multi-photon spatial qubits
haven't been relatively explored. The two-LO technique developed in
this paper will provide another tomography tools to explore spatial
qubits because it could provide the particle picture through KR
distribution and wave picture through Wigner distribution.

We would like to discuss KR and Wigner distribution separately
because the particle and wave picture of wave field can provide
independent useful information for quantum and coherent information
processing including biomedical imaging~\cite{brezinski08}.

Particle picture or local information of a wave field represented by
KR distribution has advantages in positioning or angle (momentum)
resolving to extract local activities of an target. The method can
directly locate the field or structure of an object without applying
any raw data transformation. This will be particularly useful in
cell tissue characterization such as prostate cancer cell detection.
 The KR distribution can be used to study Goos-Hanchen (GH) shifts~\cite{saleh98}
occur in near field optics and photonic waveguide. GH shifts in
position and momentum can be used to identify the loss due to
photonic crystal waveguide fabrication.

Wave picture or nonlocal information of a wave field is best
represented by Wigner function. We observe the non-positive
properties of the Wigner function for a superposition of two
spatially separated gaussian field $\mathcal{E}(x)$ analogous to a
Schr\"{o}dinger cat state in spatial coordinate. We have
demonstrated the similarities in phase space interference between
spatial properties of quantum and coherent fields via the
measurement of the Wigner function. We also show that an interesting
analogy exists between our choice of LO field and that employed in
the quantum-teleportation experiment~\cite{furusawa98}. In the x-p
representation, the small and the large beams  of  our two-LO can be
viewed as superposition of the position (in-phase) and the momentum
(out-of-phase) squeezed fields. A $\mathcal{E}(x)$ spatial gaussian
field of $TEM_{00}$ is the lowest mode and is similar to a coherent
state in phase-space picture. Gaussian beams of smaller (larger)
size than the lowest mode correspond to position (momentum) squeezed
states.  The two-LO technique here can only be used if we know the
nominal size of the signal beam. The focussed and collimated LO
beams must be chosen to achieve sufficient $x-$ and $p-$ resolution
for the given signal beam. These are the similar problems
encountered in our experiments and in the quantum teleportation
experiments which teleport an arbitrary state as a Wigner function
via EPR beams~\cite{furusawa98}.

In a complex multi-particle system or large N-biological system, we
believe KR and Wigner distributions are very useful to study the
local and nonlocal information of a macroscopic wave field such as
the mechanisms of decoherence due to neighbor particles, quantum
mapping of N-particle system or semiclassical system, and
macroscopic entanglement between two macroscopic mirrors.

In conclusion, we have demonstrated the direct measurement of KR
distribution using two-LO balanced heterodyne detection technique.
The characteristic function of KR is related to Wigner, P and Q-
distributions. Then, the Wigner, P and Q- distribution are plotted
by using raw data from the KR distribution. The physical properties
of a wave field such as local and nonlocal phase space information
are illustrated through KR and Wigner functions, respectively. This
two-LO technique can be used in information processing including
quantum information for quantum mapping and optical imaging for
biomedical applications.

\begin{acknowledgments}

\end{acknowledgments}

\end{document}